\documentclass[a4paper,10pt]{article}

\usepackage{a4wide}

\usepackage{graphicx}
\usepackage{xcolor}
\usepackage[hyphens]{url}
\usepackage{hyperref}
\hypersetup{
   colorlinks = true,
   linkcolor  = blue, 
   anchorcolor = BrickRed,
   citecolor  = blue,
   filecolor  = blue,
   menucolor  = blue,
   runcolor  = cyan,
   urlcolor  = blue
}
\usepackage[authoryear]{natbib}

\def\etal{\textit{et al.}}
\newcommand{\be}{\begin{equation}}
\newcommand{\ee}{\end{equation}}
\newcommand{\bea}{\begin{eqnarray}}
\newcommand{\eea}{\end{eqnarray}}
\newcommand{\apj}{\textit{ApJ}}
\newcommand{\mnras}{\textit{MNRAS}}
\newcommand{\prl}{\textit{PhRvL}}

\newcommand{\apjl}{\textit{ApJL}}
\newcommand{\jcap}{\textit{JCAP}}

\def\hmf{\frac{dn}{dM}}
\def\l{\left(}
\def\r{\right)}

\title{Line intensity mapping: \\ a ``novel" window to the cosmic web}

\author{Jos\'e Fonseca $^{1,2}$\\
\small $^1$Dipartimento di Fisica e Astronomia ``G. Galilei'', Universit\`{a} degli Studi di Padova,\\ \small via Marzolo 8, 35131 Padova, Italy \\
\small $^2$INFN, Sezione di Padova, via Marzolo 8, 35131, Padova, Italy\\
\small email: {\tt jfonseca@pd.infn.it}, {\tt josecarlos.s.fonseca@gmail.com}}

\date{}
\begin{document}

\maketitle

\begin{abstract}
Intensity mapping has been attracting increasing interest as a way to study galaxy evolution and the large scale structure of the Universe. Instead of detecting individual galaxies, we measure the integrated emission from a volume of the universe. Contrary to galaxy surveys, it includes light from the faintest objects in the cosmic web. We will start this overview by introducing line intensity mapping and review its current observational status. We will enumerate the most prominent emission lines for intensity mapping and discuss how to model their observed signal. The prospects of using this technique in the near future are very good with a wide range of experiments in different parts of the electromagnetic spectrum. As for any other observational tool, intensity mapping is not free of systematic uncertainties, some of which are intrinsic to it. We will briefly discuss them as well as the means to deal with foreground and background contamination. Here, we will focus on the science cases for intensity mapping in the post-reionization universe. These range from the history of star formation to measuring the acoustic scale and the fundamental physics. 

\vspace{3mm}
\noindent { \small Cosmology: large scale structure of the universe, Galaxy Evolution, Gastrophysics}

\end{abstract}

\section{Introduction}

The last century has seen great advances in our understanding of the universe, since the discovery of its expansion, as well as the Cosmic Microwave Background (CMB) and the acceleration of the universe. One of the biggest challenges of modern cosmology is to map the distribution of matter in the universe and understand how it evolved since recombination. The statistical distribution of the dark matter (DM) encodes information about the composition, evolution and initial conditions of the universe. As the universe evolved through the Dark Ages the DM over-densities slowly grew up until the Cosmic Dawn, when the first stars where born. The creation of the first stars leads to the Reionization of the universe, as the young starts in nascent galaxies emit ionising UV light. The over-densities continued growing and, after the universe becomes fully ionised around redshift 6, galaxies kept assembling, merging and forming new stars up to today.

Currently, the main way to observed the large scale structure (LSS) of the universe is by using Galaxy Surveys. Several have been done such as SDSS \citep{2013AJ....145...10D}, GAMA \citep{2011MNRAS.413..971D}, WiggleZ \citep{2010MNRAS.401.1429D}, or are currently taking data such as DES \citep{des}. Furthermore, the coming decade will see more such surveys coming online such as ESA's space mission Euclid \citep{euclid}, the optical ground-based telescopes DESI \citep{desi} and LSST\citep{lsst}, and radio galaxy surveys with SKA \citep{2015aska.confE..17A}. Galaxy Surveys have given major breakthroughs in understanding the cosmology we live in. Still, any statistically significant result requires a signal-to-noise ratio bigger than unity, i.e., the detection of thousands of galaxies ($Pn\sim 1$) in a megaparsec cubed. This limits the usage of galaxy surveys to lower redshifts. As we go higher in redshift, observed galaxies become dimmer and dimmer, its surface brightness is affected by the distance and by the acceleration of the universe. To overcome these limitations one can either increase the telescope integration time per pointing, the sensitivity, or both. The first has immediate implications on the sky area that can be covered in any reasonable amount of time. While increasing sensitivity has been the preferred path there are technological and budget limitations. An alternative is to use filters instead of spectroscopy and determine the redshift of galaxies photometrically (although photometry also suffers requires bright sources). While it is faster than spectroscopic galaxy surveys, as we go up in redshift, the angular resolution of the telescope becomes a limitation. 

In this overview, we introduce a way to map the LSS which is complementary to galaxy surveys at low redshifts. We will argue that Intensity Mapping is a compelling mean to probe the universe above redshift 3. We will start by reviewing the basic formalism underlying Intensity Mapping as well its current observational status (\S \ref{sec:overview}), followed by how to model the signal (\S \ref{sec:msig}), and current and planned experiments (\S \ref{sec:exp}). We will briefly discuss systematics and how to deal with them in \S \ref{sec:syst}. We will finish with a recollection of science cases for Intensity Mapping in \S \ref{sec:sciencecase}. 

\section{Intensity Mapping}\label{sec:overview}

\begin{figure}[!t]
\begin{center}
 \includegraphics[width=0.7\textwidth]{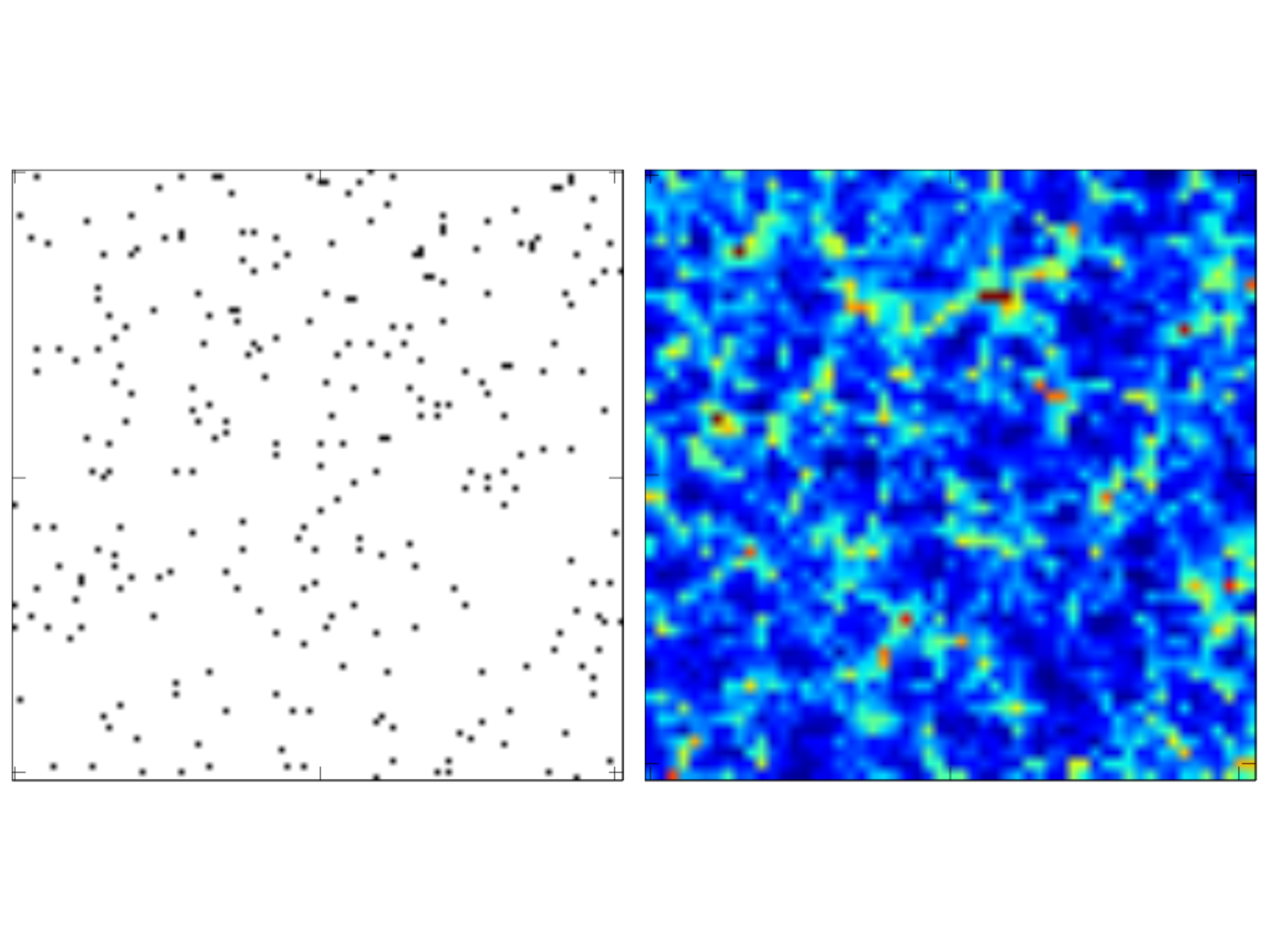} 
 \caption{Figure exemplifying how the same underlying Dark Matter field would be seen by a galaxy survey (\emph{Left}) and by an intensity mapping survey (\emph{Righ}). While a galaxy survey only detects galaxies above a certain flux threshold set by the telescope, a map of intensity gives the spatial distribution of the integrated light coming from each \emph{voxel}. (\emph{Courtesy of Marta Silva})}
 \label{fig:exampleIM}
\end{center}
\end{figure}

The easiest way to understand Intensity Mapping (IM) is by looking at Fig. \ref{fig:exampleIM}. Take \emph{a field} in the sky with a narrow redshift width. Some of the DM haloes in this field will host bright galaxies detectable by galaxy surveys. This is what is represented in the left image of Fig. \ref{fig:exampleIM}. A larger portion of the DM halos will host galaxies that are either too faint for, or still undetectable by, current telescopes. In addition, the small amount of gas in the intergalactic medium, which traces the filamentary structure of the cosmic web, is too faint for most available telescopes. But the faint galaxies host gas and dust grains that emit specific emission lines (see \S\ref{sec:msig} for details), which cumulative emission is measurable. Let us take one such line, and for now, assume that we can isolate such integrated emission from this region in the sky. This way, and with the knowledge of the emission line's rest frame wavelength, an observed frequency has a one-to-one correspondence to a redshift. Therefore an IM experiment with a given frequency and angular resolution will break this field into 3D pixels or volumes, which we call \emph{voxels}. Such an experiment produces maps like the one presented on the right image of Figure \ref{fig:exampleIM}. Instead of detecting individual galaxies, IM measures the integrated emission from physical volumes of the universe. The two images in Fig. \ref{fig:exampleIM} trace the same underlying dark matter field, while the left represents the brightest galaxies, the right image is the intensity of a given line.

IM is, therefore, a low angular resolution (and fast) way to map the large scale distribution of matter of the universe, including the contributions from faint galaxies. It is especially useful to map the clustering of the high-redshift universe. The fluctuations in the intensity statistically trace the fluctuations in the dark matter field. There are caveats though. Isolating the intensity of a target emission line is not straight forward and we will dwell into this in \S \ref{sec:syst}. While continuous emission will be removed when we look at fluctuations of the signal, contamination from interloper lines will be a systematic problem. A second caveat comes from the fact that there is intrinsic variability in the physical properties of galaxies. Despite these, on large enough scales we expect that all intrinsic fluctuations in galactic physics average out and do not affect the power spectrum of fluctuations of the emission line. Here we review the basic ideas on measuring the clustering of matter using maps of the intensity of HI \citep{2008PhRvL.100i1303C} and other lines \citep{2010JCAP...11..016V}. 

\subsection{Basic Formalism}

\begin{figure}[]
\begin{center}
 \includegraphics[width=0.65\textwidth]{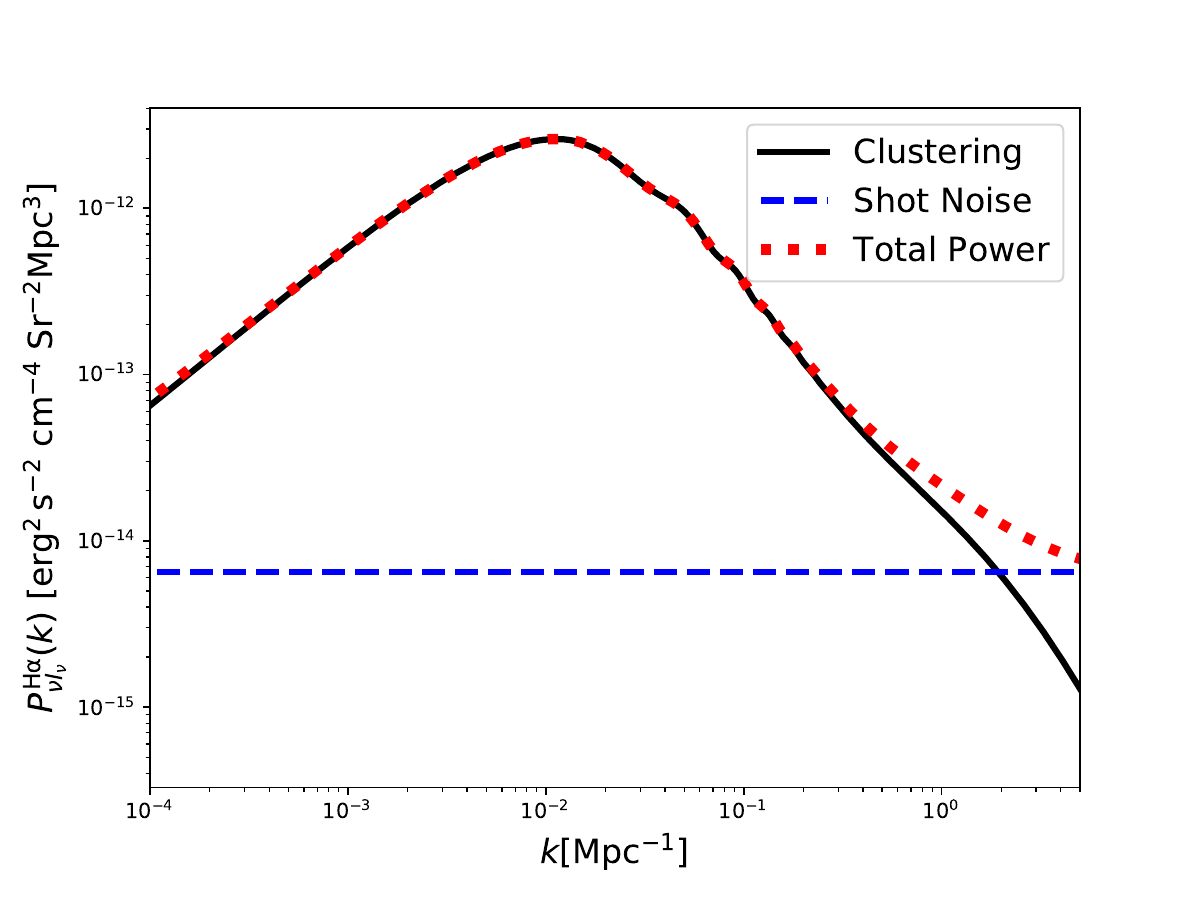} 
 \caption{3D power spectrum of H$\alpha$ IM at $z=1$ assuming \cite{2017MNRAS.464.1948F} models and Planck's fiducial cosmology. The solid black line corresponds to the clustering contribution, the dashed blue line to the shot-noise, and the total power is given by the dotted red line.}
 \label{fig:pk_Ha}
\end{center}
\end{figure}

For now let us assume we have an intensity map $I(z,\hat n)$ of a given emission line that depends on the position in the sky $\hat n$ and the redshift $ z$, and has a thickness $\Delta z$. The fluctuation map $\Delta I(z,\hat n) =I(z,\hat n)-\bar I(z)$, where $\bar I$ is the sky average intensity in the bin, traces the DM fluctuations and has the same statistical properties. Hence, the 3D power spectrum of the line intensity fluctuations at a scale $k$ is directly related to the DM power spectrum,
\begin{equation} \label{eq:pk_IM}
P_{\Delta I}(k, z)=\bar I^2(z)\ b^2(z)\ P_{\rm CDM}(k,z) + P_{\rm Shot\ Noise}(z) + P_{\rm instrumental}\,.
\end{equation}
Each term in this relation carries information from different physical effects. $\bar I$ is the average intensity of the line which depends on the physical properties of the gas and dust in galaxies. The bias of the line $b$, depends both on the emission line and the cosmology via the Halo Mass Function (HMF). The DM power spectrum $P_{\rm CDM}$ carries the bulk of the cosmological information. Similar to a galaxy survey, IM also has a shot-noise term due to the discreet nature of the emitting galaxies. In \S \ref{sec:msig} we will review how to relate the physical properties of lines with its intensity, bias, and shot-noise. The last component is the power spectrum is the gaussian error in measuring the intensity in a voxel (see \S\ref{sec:exp}). In Fig. \ref{fig:pk_Ha} we exemplify how the power spectrum of H$\alpha$ IM would look like excluding instrumental noise. In solid black we plotted the first term of Eq. \ref{eq:pk_IM}, in dashed blue the shot-noise and in dotted red the total power. We see that on very small scales the shot-noise can dominate over the clustering.

\subsection{Observational Status}

\begin{figure}[]
\begin{center}
 \includegraphics[width=0.7\textwidth]{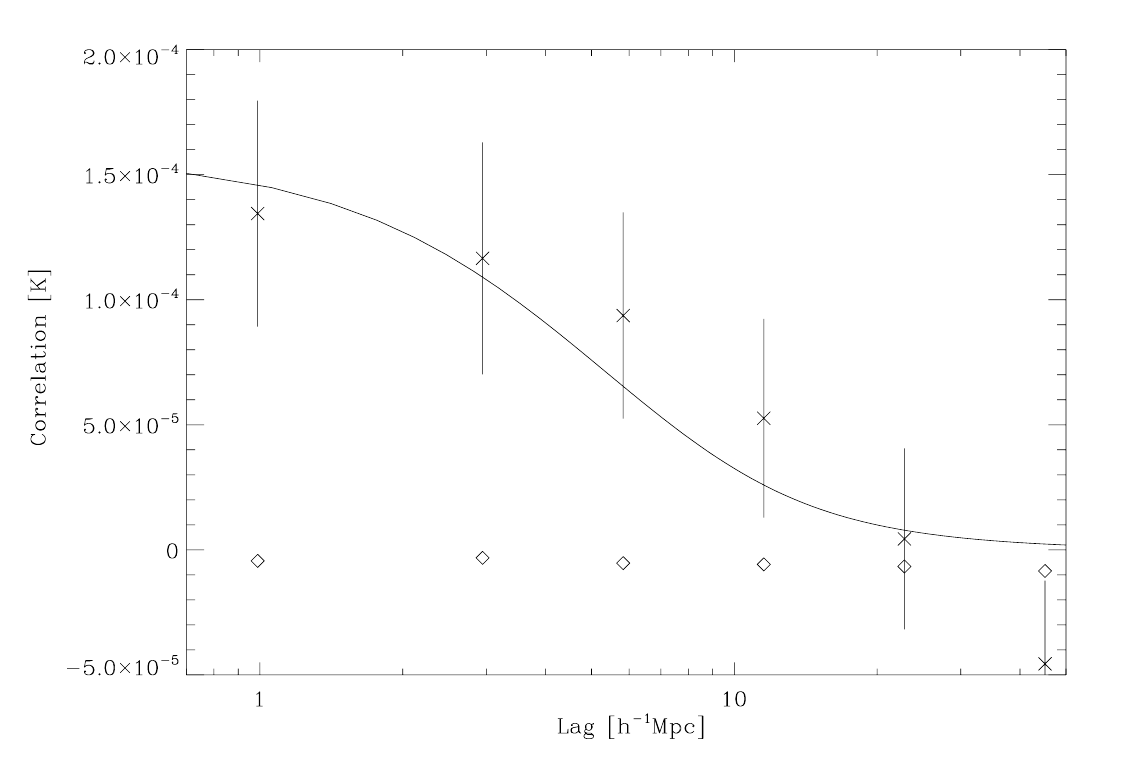} 
 \caption{Cross-correlation between the 21cm brightness temperature and the DEEP2 galaxy survey \citep{2010arXiv1007.3709C}. \textit{Figure reproduced with permission, cortesy of Tzu-Ching Chang}.}
 \label{fig:hiim_1st_detect}
\end{center}
\end{figure}

One may think that IM is still just a proposal. In fact, there have been several observational studies, including detections. The first ever detection of the HI signal was done by \cite{2010arXiv1007.3709C}. They used the Green Bank Telescope (GBT) to look for the 21cm emission from the redshift range of $0.53 < z < 1.12$, centred at $z\simeq0.8$. The targeted fields overlapped with part of the DEEP2 survey \citep{deep2}, an optical redshift survey done with the Keck II telescope. Fig. \ref{fig:hiim_1st_detect} reproduces their results and shows the measured correlation between the 21cm brightness temperature with the DEEP2 galaxy survey in crosses. The solid line corresponds to the expected correlation between the two samples. The diamonds in the figure represent a null-test with a 1000 random realisations of the galaxies position. This result was a major breakthrough, not only was the first detection of the cross-correlation but more importantly, it has shown that neutral hydrogen does trace the underlying DM field. It confirms that similar to galaxy surveys, one can use the intensity of a line to map the large scale distribution of matter. The result worked as a proof of concept and has lead to further studies with the GBT. \cite{2013MNRAS.434L..46S} increased the covered area to $\sim 41\deg^2$ and integrated over 190 hours, and cross-correlated the HI signal with the WiggleZ galaxy survey \citep{2010MNRAS.401.1429D}. Using both the auto and the cross-correlations they were able to constraint the HI bias and abundance, $\Omega_{\rm HI} b_{\rm HI}= 0.62^{+0.23}_{-0.15} \times 10^{-3}$. 

There have been more attempts to detect line IM beyond HI. \cite{2016MNRAS.457.3541C} cross-correlated Lyman-$\alpha$ emission with Quasars at redshift 3, both obtained from SDSS-III \citep{2013AJ....145...10D}. To obtain the Ly$\alpha$ signal they took the DR12 SDSS's Luminous Red Galaxy (LRG) sample and removed the best fit models for the LRGs spectra. The residual light should be Ly$\alpha$ emission from $z\sim3$. Although their result was inconclusive, it shows that current data sets may already be used to obtain maps of intensity. Another example of tentative detections of line IM in cross-correlation is the work of \cite{2018MNRAS.478.1911P} with a higher success. The authors cross-correlated some of Planck's bands \citep{2010AA...520A...9L} with DR12 SDSS quasar sample and SDSS's low redshift CMASS galaxies, and measured the [CII] emission from $z\sim2.6$, the major component of the Cosmic Infrared Background (CIB). Not all searches have been done in cross-correlation. The CO Power Spectrum Survey (COPSS) used the Sunyaev-Zel'dovich Array over 0.7$\deg^2$ and has detected the power spectrum of CO(1-0) fluctuations at $z\simeq3$, with a 3$\sigma$ detection of the signal \citep{2016ApJ...830...34K}.

All these attempts have shown that Line Intensity Mapping is indeed a tool to probe the cosmos. Most surveys have been done is over very small sky areas, except \cite{{2018MNRAS.478.1911P}} with an overlap area of 6483$\deg^2$. None of the used experiments were optimised for IM, but still produced tentative detections of the signal. Fundamentally they have set the bases for future experiments targeting a line or sets of lines. 

\section{Modeling the signal} \label{sec:msig}

Prior to modelling the luminosity of a line with respect to its host galaxy properties, one needs to identify which lines can potentially be used in intensity mapping probes. In principle, any strong galactic emission line is a good candidate. Still, any such line should be consistently emitted by galaxies of the same type. This excludes strong lines from quasars as other galaxies may not emit them, and strongly favours lines associated with star-forming galaxies. Furthermore, the astrophysical processes creating them are fairly known. Young blue stars in these galaxies emitted UV radiation ionising or exciting neutral hydrogen, which then cascades down emitting lines such as Ly$\alpha$, H$\alpha$ or H$\beta$. In the UV/optical part of the spectrum, we can also find very strong metal lines such as [OII] 373nm, the [OIII] doublet (495.9 nm \& 500.7 nm), Nitrogen and Sulphur lines. Most of these lines are typically used to measure the redshift of Emission Line Galaxies (ELGs). But there are other lines present in star-forming galaxies. In the Far-Infrared (FIR), [CII] (157$\mu$m) and the rotation lines of CO are an efficient cooling mechanism in $H_2$ clouds that lead to star formation. Dust in galaxies works as re-processing bolometers emitting in the mid-range of the infrared. These range from all sorts of emission lines from ionised Oxygen, Nitrogen, Neon, Silica, Sulphur. In addition, any DM halo with neutral hydrogen will emit in the radio band the 21cm line, also referred to as HI. This hyperfine spin-flip transition stands alone against continuum radio emission from galaxies, being a major line in current IM studies. 

To predict the luminosity of a given line one needs to take into account all the physics present in the Interstellar Medium (ISM) which may vary between galaxies and within a galaxy. The crucial assumption in IM is that, if the voxel we consider is big enough, therefore containing several galaxies, then such scatter averages out. From this, it follows that one can model the average emission of a line using the macroscopic properties of galaxies. Generally one finds two modelling approaches. One is empirical and uses scaling relations between the line luminosity and other properties of a galaxy. Despite being the easiest, fastest and most straight forward manner to estimate the signal, it relies on calibrations of the scaling relationships done in limited ranges of halo mass, luminosity, metallically and redshift. Alternatively, one can use numerical simulations and semi-analytical models to compute the rate of recombinations from radiative processes and collisional excitations in the gas, as well as counting photons that escape the galaxy. These can take into account many more details of the ISM physics and galaxy evolution but, such models have lots of free parameters and requires several assumptions, as well as being computationally demanding. An example of this approach is the usage of hydrodynamical simulations to obtain the HI signal \citep{2015ApJ...814..146V}. Another example is the recent work of \cite{2019MNRAS.482.4906P}, assessing the impact of different assumptions in numerical simulations over the estimated signal of FIR carbon lines (CO's, [CII] and [CI]). 

Here we will review the principles of the empirical approach, as they are a simple and tangible introduction to the modelling involved. In a typical galaxy, the star formation rate (SFR) is the energy source for other emission lines, hence we write the relationship
\be \label{eq:LSFR}
L_{int} =R \times \l \frac{SFR \l M,z\r}{\rm M_\odot /yr}\r \,.
\ee
This way one assumes that the intrinsic luminosity $L_{int} $ of a line or part of the spectrum is linearly dependent on the SFR which, in turn, depends on the halo mass $M$ of the host galaxy. This relation is valid for the UV emission, Ly$\alpha$, H$\alpha$ and H$\beta$, UV and optical oxygen emission lines, and for the FIR emission. As an example, one can compute the value of $R$ for some lines if we assume an optically thick interstellar medium, case B recombinations, a Salpeter universal initial mass function and that all the UV continuum has been absorbed by the gas in the galaxy. For more details look at \cite{2017MNRAS.464.1948F} and references therein. Note as well that the intrinsic luminosity still needs to be corrected by the extinction $E$, as part of the emission will be absorbed by dust and/or gas in the galaxy. This means that 
\be \label{eq:lobs}
L = 10^{-E(\nu)/2.5} L_{int}\,.
\ee
The extinction depends on the part of the spectrum we are considering and may depend as well in the halo mass. Indeed one expects lower mass halos to have fewer metals. The case of the FIR lines is slightly more complicated, but generally one assumes that they depend on the total FIR luminosity, which we parametrise as
\be \label{eq:l}
\log_{10}\l L_{line}\r=\alpha\log_{10}\l L_{FIR}[{\rm L_{\odot}}]\r+\beta\,.
\ee
The pair $(\alpha,\beta)$ are measured observationally for the line in consideration, although currently there is still a high level of uncertainty on the fits. It is also plausible to assume that they depend on the redshift, halo mass and/or metallically. \cite{2017MNRAS.464.1948F} summarised currently available calibrations for several FIR emission lines. 

\begin{figure}[!ht]
\begin{center}
 \includegraphics[width=0.49\textwidth]{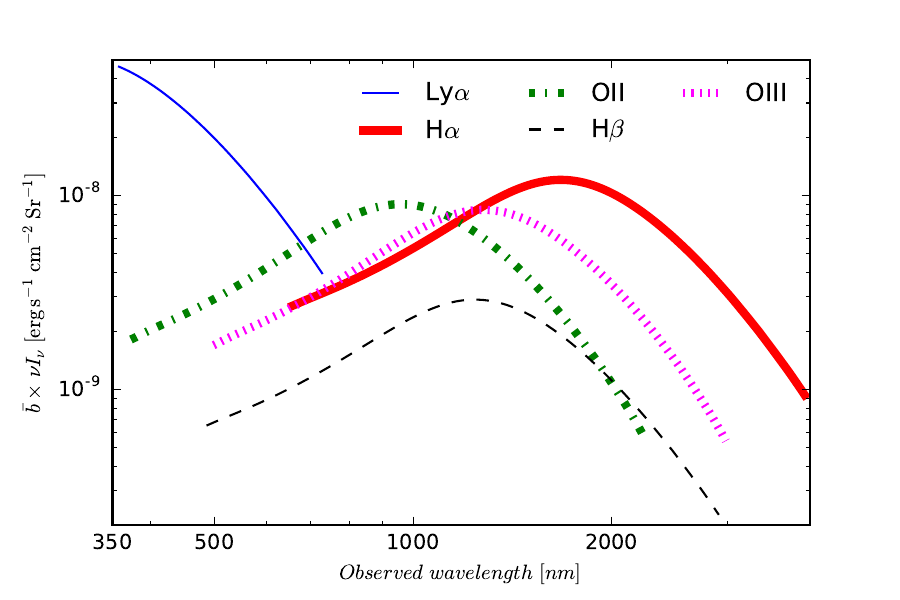} 
 \includegraphics[width=0.49\textwidth]{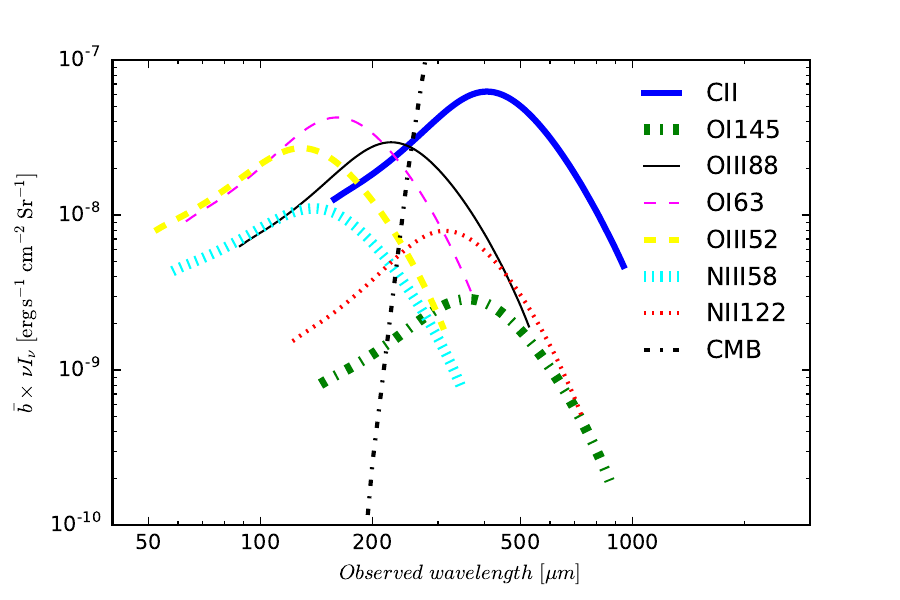} 
 \includegraphics[width=0.49\textwidth]{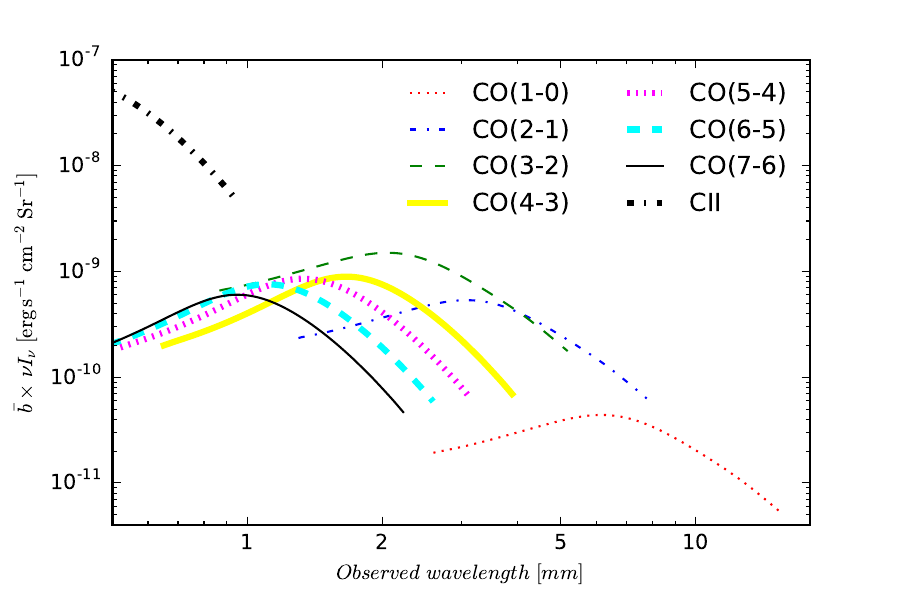} 
 \caption{Estimates of the product $b\times \nu I_\nu$ as a function of the observed wavelength for several galactic emission lines. \emph{Figure reproduced from} \cite{2017MNRAS.464.1948F}.}
 \label{fig:bnuI}
\end{center}
\end{figure}

Once we have determined the line luminosity, one computes the average intensity from a volume of the universe weighted by the comoving HMF \citep{Sheth:1999mn}. Then, the average observed intensity of a given emission line is given by
\be \label{eq:savehm}
\bar I_\nu (z)=\int_{M_{min}}^{M_{max}} {\rm dM}~\hmf ~\frac{L(M,z)}{4\pi D_L^2}~\tilde y D_{A}^2 \,.
\ee
Note that the HMF $dn/dM$ has units of Halos/Mpc$^3$/$M_\odot$, $D_L$ is the luminosity distance, $D_A$ is the comoving angular diameter distance and $\tilde y \equiv d\chi/d\nu=c\l 1+z\r^2/\nu_{line}H\l z\r$. Equivalently, one can re-write Eq. \ref{eq:savehm} in terms of luminosity where the HMF becomes the line luminosity function. We can now see the importance of relating the line luminosity to the halo mass, as any realistic estimate of the line intensity needs to take into amount the number density of galaxies of different masses. This formalism also allows us to compute the bias of an emission line using the halo bias $b_h$
\be
\label{eq:lumbias}
b \l z\r \equiv\frac{\int^{M_{max}}_{M_{min}}{\rm dM} ~b_h\l M,z\r L(M,z) ~\hmf }{\int^{M{max}}_{M_{min}}{\rm dM}~ L(M,z) ~\hmf }\,,
\ee
and its shot-noise
\be \label{eq:pshotline}
P_{Shot\ Noise} (z)=\int_{M_{min}}^{M_{max}} {\rm dM}~\hmf \l\frac{L(M,z)}{4\pi D_L^2}~\tilde y D_{A}^2\r^2 \,.
\ee
This formalism requires that we know the halo mass range of galaxies producing the line under consideration. While it is reasonable to assume that most halos have hydrogen, metals lines may only be present in intermediate-to-high mass halos, which had more time for metal enrichment of the ISM. In Fig. \ref{fig:bnuI} we exemplify how these models translate into estimates of the signal and on how the strength of the lines compare with each other. 

Modelling HI is slightly different from the other discussed so far. In thermal equilibrium, the environment's temperature sets the probability of an atom be in an excited state (with the spin aligned) or in the ground state (with the spins in opposite directions). We can divide the neutral hydrogen number density into contributions from the two states $n_{HI}=n_0+n_1$, with its ratio given by
\be
n_1=n_0\frac{g_1}{g_0}e^{-\frac{h\nu_{21}}{k_B T}}\,,
\ee
with $g_1=3$, $g_0=1$. We expect the environment temperature to be much bigger than the temperature of the line, so $n_1\simeq3/4 n_{HI}$. Hence the emissivity of the line (thus the luminosity and flux) will depend on the amount of neutral hydrogen in the galaxy. Therefore, cosmology with HI IM only requires us to understand how the amount of hydrogen relates with the halo mass, $M_{HI}(M)$, of the host galaxy. \cite{2015aska.confE..19S} used the power law $M_{HI}=AM^\alpha$, finding $\alpha\sim 0.6$ to be a good match for the low and higher redshift data available, and calibrating $A$ with the measurements of \cite{2013MNRAS.434L..46S}. As before, we relate the temperature of the signal, bias and shot noise with respect to the halo model. The temperature is given by
\be
\bar T_{HI}(z) \approx 566h \l\frac{H_0}{H(z)} \r \l\frac{\Omega_{HI}(z)}{0.003} \r \l1+z\r^2\, \mu K\,,
\ee
where $\Omega_{HI}(z)\equiv \l1 + z\r^{-3}\rho_{HI}(z)/\rho_{c,0}$, and the HI energy density is given by
\be
\rho_{HI}(z)= \int_{M_{min}}^{M_{max}} {\rm dM}~\hmf M_{HI}(M) \,.
\ee
The bias and shot noise are then given by 
\bea
b_{HI}(z)&=& \rho_{HI}^{-1} \int_{M_{min}}^{M_{max}} {\rm dM}~\hmf M_{HI}(M) b(M,z) \,,\\
P_{HI}^{shot}(z)&=& \l\frac{T_{HI}(z)}{\rho_{HI}} \r^2\int_{M_{min}}^{M_{max}} {\rm dM}~\hmf M^2_{HI}(M) \,.
\eea
For more details on the derivation of the HI temperature and how to estimate the halo mass range that has neutral hydrogen, please look at \cite{2015ApJ...803...21B}.

\section{Experimental Landscape} \label{sec:exp}

Any IM experiment has to be able to pixelise the sky and take spectra of each pixel. The voxel is then set by the angular $\delta\theta$ and spectral $\delta\nu$ resolutions of the experiment. Note that Intensity Mapping is by default a low angular resolution technique, hence the technical requirements are spectroscopy of fields rather than point sources. The most straight forward manner of doing it is by using integral field units (IFUs) but these are costly and may take a long integration time for a large survey. Alternatively one can do imaging of the sky with a CCD and, narrow band or Fabry-P\'erot filters in the instrument optics. These solutions are for the optical and NIR part of the spectrum. While some may be still applicable in the far-infrared, for the radio and submillimetre one requires other solutions. In general, one would have to put a calorimeter, spectrograph, bolometer or receiver at the focal plane of an antenna or telescope. Further down into the radio frequencies one needs dipoles to measure any HI emission from the high redshift universe. From the submillimetre into the low frequencies one can use the telescopes or antennas in single-dish mode, or as interferometers to gain angular resolution. In fact, this opens a wide range of experimental possibilities from the ground and from space, including balloon experiments on the upper atmosphere. Still, one should note that the UV and parts of the infrared can only be seen from space. Ground-based experiments at higher altitudes can observe the FIR part of the spectrum. The altitude requirement can be relaxed as we move into the radio frequencies. Although a lot of optical lines are interesting candidates for IM study, they quickly redshift to the NIR where the atmospheric observational window becomes more opaque.

At the moment only a handful set of experiments have been designed as intensity mappers, mainly targeting the Epoch of Reionisation (EoR). Other experiments have been designed with diverse science goals and techniques in mind, but their experimental settings allow for IM surveys. Indeed this is the case of MeerKAT, SKA-MID or HETDEX, which now have IM surveys proposed. In table \ref{tab:experiments} we list current, planned, or proposed experiments targeting (or with the potential to target) the post-EoR universe using intensity mapping of one or more lines. We understand \emph{planned} as funded or partly funded. Note that when a line extends well into the EoR we left redshift range open-ended, as some may have low abundances at those redshift ranges. 

\begin{table}[!ht]
 \begin{center}
 \caption{List of IM experiments which can probe the LSS of the post-EoR universe.}
 \label{tab:experiments}
 {\scriptsize
 \begin{tabular}{|l|c|c|c|c|c|c|}\hline 
{\bf Experiment} & {\bf Line} & {\bf Frequency } & {\bf Redshift } & {\bf Type of } & {\bf Location} & {\bf State} \\ 
& & {\bf Range} & {\bf Range} & {\bf experiment} & & \\ 
 \hline
 SKA-LOW & HI & 0.05-0.35 & $>$3 & Dipole & Australia &planned \\ 
\cite{2015aska.confE...1K} & & GHz& & Interferometer & & \\ \hline
SKA-MID & HI & 0.35-15.8 & 0-3 & Single Dish & South &planned \\
\cite{2015aska.confE..19S}& & GHz& & Array & Africa & \\ \hline
CHIME & HI & 0.4-0.8& 0.78-2.5 & Cylinder & Canada &running \\
\cite{2014SPIE.9145E..22B}& & GHz& & Interferometer & & \\ \hline
HIRAX & HI & 0.4-0.8 & 0.78-2.5 & Dish & South &planned \\
\cite{2016SPIE.9906E..5XN}& & GHz& & Interferometer & Africa & \\ \hline
Tianli & HI & 0.54-1.5 & 0-2.5 & Cylinder & China &running \\
\cite{tianli}& & GHz& & Interferometer &  & \\ \hline
MeerKAT & HI & 0.58-1.67 & 0-1.45 & Single Dish & South &running \\
\cite{meerklass}& & GHz& & Array & Africa & \\ \hline
BINGO & HI & 0.96-1.26 & 0.13-0.48 & 50 feedhorns & Brazil &proposed \\
\cite{2018arXiv180301644W}& &GHz & & telescope &  & \\ \hline
COPSS & CO(1-0) & 27-35 & 2.3-3.3 & Dish  & USA &finished \\
\cite{2016ApJ...830...34K}& & GHz& & Interferometer &  & \\ \hline
COMAP & CO(1-0) & 30-34 & 2.4-2.8 & Spectrometer  & USA &running \\
\cite{2016ApJ...817..169L}& & GHz & & Array &  & \\ \hline
AIM-CO & CO(2-1) & 86-102 & 1.2-1.5 & Dish & USA &proposed \\
\cite{aimco}& CO(3-2)& GHz & 2.4-3& Interferometer &  & \\ \hline
CONCERTO & [CII] 157$\mu$m & 125-360 & 4.3-14.2 & Spectrometer & Chile & planned \\
\cite{2018AA...609A.130L}& CO(4-3) & GHz& 0.28-2.68& Array &  & \\ \hline
CCAT-prime & [CII] 157$\mu$m & 185-440 & $>$3.3 & Telescope & Chile &planned \\
\cite{2018SPIE10700E..5XP}& & GHZ & & &  & \\ \hline
TIME & [CII] 157$\mu$m & 200-300 & $>$5.3 & Spectrometer & USA &running \\
\cite{timep}& CO(4-3) & GHz& 0.53-1.3& Array &  & \\ 
& CO(3-2)& & 0.15-0.73 &  &  & \\ \hline
 Origins & [CII] 157$\mu$m & 0.45-50 & 0-3.2 & Spectrograph & Space & proposed\\ 
\cite{2018arXiv180909702T} & [NII] 122$\mu$m & THz& 0-4.4 & & &\\ 
& [OIII] 88$\mu$m & & 0-6.5 & & &\\ \hline
STARFIRE& [CII] 157$\mu$m & 0.71-1.25 & 0.5-1.7 & IFU & Balloon & planned \\
 \cite{starfire} & [NII] 122$\mu$m  & THz & 0.97-2.5 & &&\\ \hline
 CDIM& H$\alpha$ & 40-400 & $>$0.14 & Linear  & Space & proposed \\
 \cite{2016arXiv160205178C} & [OIII] 500nm & THz& $>$0.5 &Variable &&\\ 
  & [OII] 373nm & & $>$1.0 & Filters&&\\ \hline
 SPHEREx& H$\alpha$ & 60-400 & 0.14-6.6 & Narrow & Space & planned \\
 \cite{2014arXiv1412.4872D} & [OIII] 500nm & THz & $>$0.5 & Band&&\\
 & [OII] 373nm & & $>$1.0 &Filters &&\\ \hline
 HETDEX & Ly$\alpha$ & 550-854 & 1.9 - 3.5 & IFU & USA & running\\ 
 \cite{hetdex}& & THz& &  & & \\ \hline
 MESSIER& Ly$\alpha$ & 1500 & z=0.64 & Narrow band & Space & proposed\\ 
\cite{{2017ApOpt..56.8639M}} &&THz&& filter& &\\ \hline
 \end{tabular}
 }
 \end{center}
\vspace{1mm}
\end{table}

Each experiment will have its own sensitivity $\sigma_N$ which depends on the detector, the integration time and the optical setup. The instrumental noise is usually taken to be gaussian centred at zero. This means that its two-point function is simply given by the variance on the instrumental noise. In turn, this translates into an instrumental noise power spectrum which contributes to the total 3D power spectrum is given by 
\be \label{eq:pnoise}
P_{instrumental}=\sigma_N^2(z)\times V_{pixel}(z)\,.
\ee
Note that $\sigma_N$ normally depends on the observed frequency, which for a given line, is uniquely dependent on the redshift. The pixel volume $V_{pixel}$ can be written as 
\be
V_{pixel}=\frac{c\l1+z\r^2\chi^2(z)}{\nu_{line}H(z)} \delta\nu \delta\theta^2\,,
\ee
and depends both on the target line and the redshift of interest, as well as in the angular and frequency resolution of the experiment. Although Eq. \ref{eq:pnoise} seems to indicate that to improve results one either increases the sensitivity per pixel, or the resolution, this is not necessarily true, as the sensitivity is dependent on the resolution of the experiment. 

\section{Systematics} \label{sec:syst}

Any experiment has to deal with systematics. Indeed the least astute reader can think of one or two issues that may arise when obtaining the signal from any realistic experimental setting. An exhaustive exploration of how to deal with systematics is beyond the scope of this overview. But it is import to identify the major contaminants and how to deal with them, especially the ones that are intrinsic to IM. In general, one can divide the contaminants into 3 groups: instrumental, continuum emission and interloper lines. Instrumental effects may concern with calibration, mode mixing caused by the frequency dependence of the beam or the instrumental optics, non-stable noise terms (such as the $1/f$ noise in radio receivers), etc. They are dependent on the experimental design which varies wildly between parts of the electromagnetic spectrum. Contamination from continuum sources and interloper lines is specific to IM, therefore we will focus on those. 

We understand continuum emission as any foreground or background source of light with a smooth dependence in frequency and some anisotropic angular structure in the sky. If a foreground is totally isotropic (for small patches of the sky this is a good assumption) then it doesn't affect the fluctuation map, as the removal of the average intensity removes any isotropic contamination (although it may contribute to the error budget). While in the radio frequencies we don't expect a radio background, this is not true in the FIR with the CIB or the CMB, or the Optical and near-IR with AGN and continuum stellar UV emission from high redshift. At lower frequencies, continuous foregrounds include synchrotron emission from the galaxy, galactic and extragalactic free-free emission and point sources. The milimetre and infrared see continuous thermal dust emission from the galaxy. From the NIR down to the UV continuous foregrounds are not so well known. But we can use our knowledge of the spectral smoothness of continuous contaminants to remove them. \cite{2012ApJ...756...92C} estimated the NIR background light which can then be fitted out from intensity maps. Alternatively, one can also use blind methods (as Principle Component Analyses or Independent Component Analyses) to remove residual foregrounds and reconstruct the power spectrum \cite[see e.g.,][for HI]{2015MNRAS.447..400A}.

Interloping lines are a contaminant unique to IM. An interloper is an emission line different from the targeted one, which comes from another redshift but is observed in the same desired frequency. To exemplify the problem let us assume we are observing at $1.97\mu$m targeting H$\alpha$ emission from $z=2$. In the same wavelength one receives contributions from [OII] at $z=4.3$, H$\beta$ at $z=3.1$, [OII] at $z=2.9$ and [SII] (671.7nm) at $z=1.9$. So interlopers are a serious source of confusion with which we have to deal. The most straight forward prescription is to mask pixels. If we know which one is the interloper line, then we can use galaxy catalogs for targeted masking. Otherwise, one can blindly mask pixels above a certain flux threshold. Both methods suppress the power spectrum of the interloper but also of the target line. Masking is more appropriate for interlopers with emission stronger or comparable to the target lines, or when the interloper comes from a redshift considerably lower than the target one. Pixel masking may not be suitable for low angular resolution experiments, as it will reduce substantially the signal-to-noise ratio. Another way of dealing with interlopers is by using their clustering properties. The interloper is also a tracer of the LSS but at a different redshift. Therefore the observed power spectrum will have a contribution from the target line and a further contribution. This one is the power spectrum of the interloper at its emission redshift projected onto the redshift of the target line \citep{2014ApJ...785...72G}. \cite{2016ApJ...832..165C} use this anisotropic power spectrum separation method in an MCMC to distinguish [CII] and CO emission from a mock data cube, given that the survey is sensitive enough. 

Finally one can deal with both interloper lines and continuum emission using cross-correlations, as done in obtaining the first IM detections. \cite{2019ApJ...872...82S} have used this, together with the fact that continuum foregrounds are correlated in frequency to do component separation and obtain maps of [CII] emission. As future surveys will do tomography of large volumes of the universe, these types of approaches become important to separate the signal of several lines and continuum emission from different sources. Although this implies substantial amounts of forward modelling and computational power, it is the most reliable as we would do component separation and parameter estimation at the likelihood level using bayesian statistics. 

\section{Science Cases} \label{sec:sciencecase}

\begin{figure}[!t]
\begin{center}
 \includegraphics[width=0.5\textwidth]{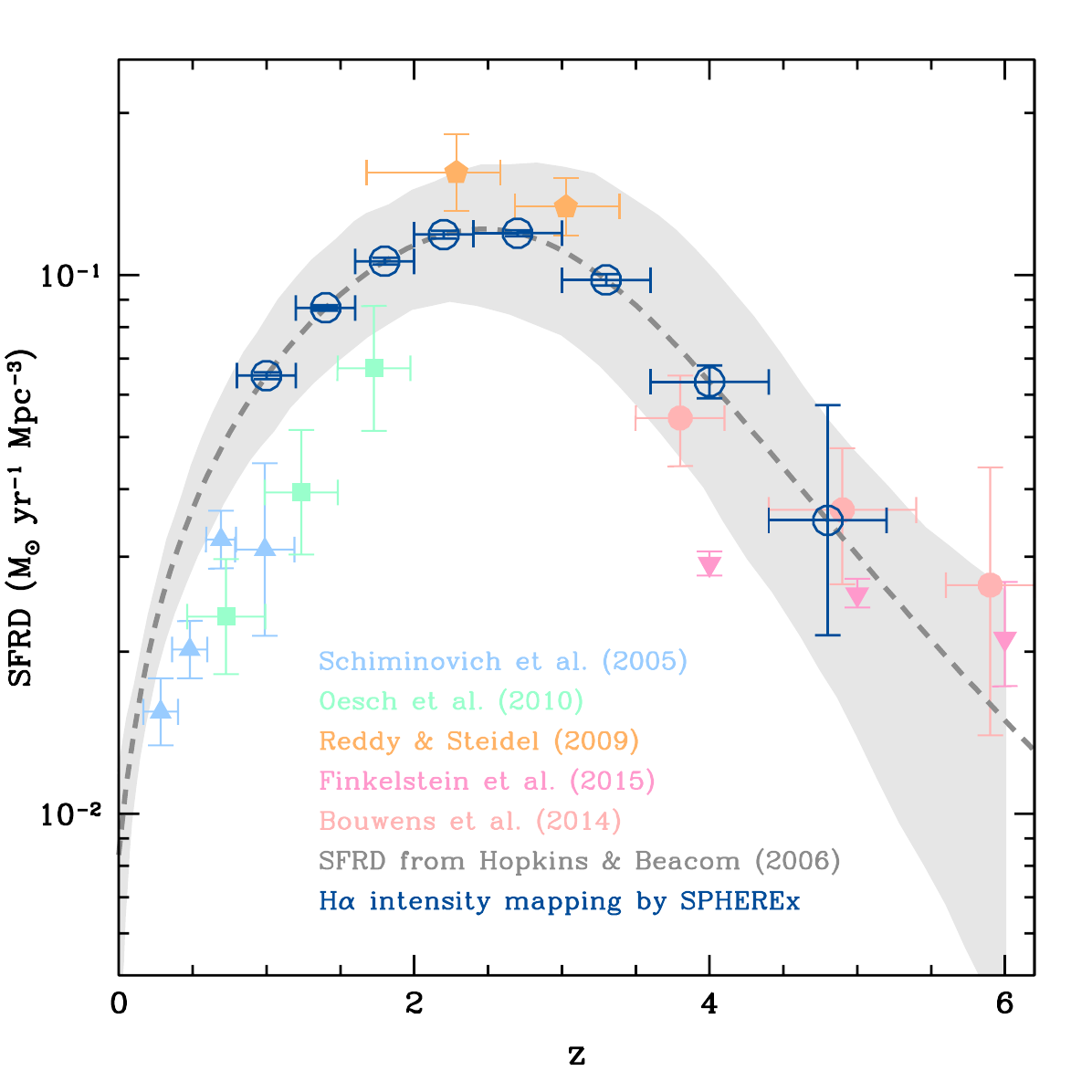} 
 \caption{Forecasted constrains on the star formation rate density using H$\alpha$ IM with SPHREx. \emph{Figure reproduced with permission, from} \citet{2017ApJ...835..273G}.}
 \label{fig:sfrd_constrains}
\end{center}
\end{figure}

To complete this short overview of the field of intensity mapping we need to scan through its research potential. From what the reader can see the fact that we do not need to detect individual galaxies means that we can push our studies to higher and higher redshifts, and wider parts of the sky. Therefore we can study the Large Scale Structure of the universe up to the end of the EoR. One can use IM to measure cosmological parameters in a similar manner to galaxy surveys. But, as we saw, the IM signal entangles cosmology with properties of the gas, the host galaxy, and galaxy evolution. Therefore, not only we can test $\Lambda$CDM (and non-$\Lambda$CDM as we will see), one can constrain line emission models, the fainter end of line luminosity functions and the star formation history of the universe (see for example Fig. \ref{fig:sfrd_constrains}). One can divide the science cases into two: Cosmology and Galaxy Evolution.

A lot of effort has been put in studying how the proposed IM surveys will measure the standard cosmological parameters such as $\sigma_8$, the growth $f$, the Dark Energy equation of state ($w_0$,$w_a$), $H(z)$, $D_A(z)$ and the BAO scale, as well as the matter power spectrum. \cite{2015ApJ...803...21B} have extensively studied the constraints on these parameters from 21 cm IM experiments in the late universe, and subsequently updated by \cite{Pourtsidou:2016dzn} with the inclusion of correlation with optical galaxy surveys and the CMB. Although these forecasts have been done for experiments using HI IM one can use other lines. As an example, SPHEREx can measure the power spectrum of H$\alpha$, H$\beta$, [OIII] and [OII] IM with varying SNR \citep{2017ApJ...835..273G}. \cite{2017MNRAS.464.1948F} has looked at the detectability of the power spectrum using these UV/optical lines as well as the CO rotation lines and the [CII] forbidden line, concluding that all are competitive probes of the large scale structure of the universe at $z\sim2$, when galaxy surveys can no longer provide large enough samples. 

These studies looked for constraints of standard $\Lambda$CDM cosmology. In fact, one of the most compelling science cases for IM is all the beyond $\Lambda$CDM science that it can do. The reason lies in the technical characteristics of IM. On one hand, it can provide information very deep in redshift and therefore test the growth history, as well as DE models and gravitational theories. \cite{2016ApJ...817...26B} has studied how HI IM with SKA-MID will constrain these models for different modified gravity parametrizations. On the other hand, IM can provide fast large sky surveys, which combined with the redshift depth, probes volumes so big that it allows us to tap into physical effects only seen on super-horizon scales. This is the case of primordial non-Gaussinaity ($f_{\rm NL}$) of the local type that induces a scale dependence on the bias of all DM tracers, which peaks on the very large scales. While \cite{2013PhRvL.111q1302C} looked at the constrains produced by HI IM using the SKA-MID, \cite{2019ApJ...872..126M} studied the effect of primordial non-Gaussianity in 2 and 3-point statistics of [CII] and CO(1-0) IM. In addition to $f_{\rm NL}$, the so-called ``GR effects", leave imprints on the observed power spectrum that can only be observed on the largest cosmological scales \cite[e.g.,][with HI IM]{2015ApJ...814..145A}. These IM surveys are also cosmic variance limited as galaxy surveys are. These results can be improved with cross-correlations between maps of intensity and galaxy surveys \citep[e.g.,][]{2015ApJ...812L..22F}, or the CMB \citep[e.g.,][]{2019MNRAS.485.1339B}, or even other between several maps of intensity \citep[][for HI and H$\alpha$ IM]{2018MNRAS.479.3490F}. Besides the large scale effects, \cite{2015ApJ...814..146V} used hydrodynamical simulations to study the effect of massive neutrinos on the clustering of neutral hydrogen, forecasting that SKA-MID together with SKA-LOW can put a 2$\sigma$ constraint on the sum of the neutrino masses.

\begin{figure}[!t]
\begin{center}
 \includegraphics[width=0.7\textwidth]{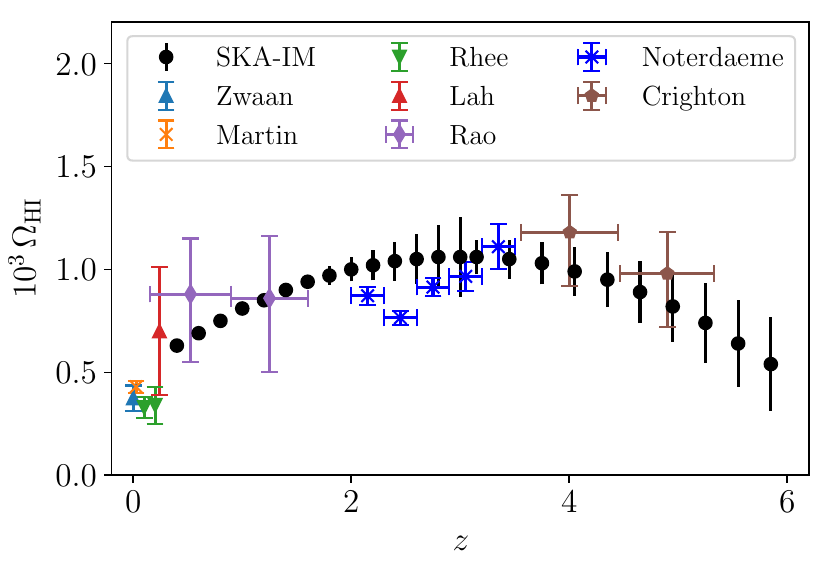} 
 \caption{Forecasted constrains on the amount of neutral Hydrogen using 21cm IM with SKA-MID and SKA-LOW. \emph{Figure reproduced with permission from} \citet{Pourtsidou:2016dzn}, \emph{courtesy of Alkistis Pourtsidou}.}
 \label{fig:OHI_constrains}
\end{center}
\end{figure}

But understanding the properties of galaxies and/or the gas and their evolution is also a major driver of IM, if not the major one, especially because one cannot detect large numbers of galaxies higher redshift. As an example of the potential of IM, \citet{2017ApJ...835..273G} used the specifications of SPHEREx and the fact that it will perform H$\alpha$ IM, to determine how such experiment would constraint the SFR density (SFRD). One can see their forecasted conditional errors in Figure \ref{fig:sfrd_constrains}. Even though the SFRD is degenerate with other amplitude parameters, such as the bias or $\sigma_8$, one can see that SPHEREx will produce very competitive constrains on the star formation history. The results in Fig. \ref{fig:sfrd_constrains} only refer to H$\alpha$ and should be improved once including other lines such as H$\beta$, [OIII] and [OII]. Another example of galactic properties one can infer using IM is the amount and redshift evolution of neutral hydrogen, as shown in Fig. \ref{fig:OHI_constrains}. \citet{Pourtsidou:2016dzn} used the 3D power spectrum of HI from SKA-MID (up to $z=3$) and SKA-LOW (for $z>3$) to compute the constrains on the HI density across cosmic times. But the HI density can also be probed using the shot noise term of HI IM cross-correlated with optical galaxies, as shown by \cite{Wolz:2017rlw}. Any sort of realistic parameter estimation from maps of intensity needs to do a joint likelihood estimation including cosmological and astrophysical parameters, as proposed in \cite{2017ApJ...848...23K} for EoR experiments.

Besides this, IM detects even the contribution of the faintest sources in the cosmic web in its signal. This makes it a unique mean to trace the Inter-Galactic Medium and the filaments. \cite{2016MNRAS.462.1961S} used Ly$\alpha$ IM to study how next-generation space-based UV telescopes, with high signal-to-noise ratio, can distinguish between galactic Ly$\alpha$ emission and the one from the filaments.

\section{Summary}

We hope we have convinced the reader that IM is a powerful new tool to map the large scale distribution of matter. We went from detecting individual sources such as galaxies to measuring the integrated emission of a region of the universe. Then spacial fluctuations between voxel's intensity are due to fluctuation in the underlying dark matter. For Cosmology, IM will be complementary to galaxy surveys in measuring cosmological parameters. Its competitiveness comes from the fact that it can map large volumes of the universe faster than conventional galaxy survey, but still keeping a good redshift resolution. Besides, it will provide information on galaxy evolution up to the EoR which conventional galaxy surveys are unable to do. We will also be able to learn about the average macroscopic emission properties of the gas and metals in galaxies. Furthermore, it will allow us to improve our understating of the faintest elements of the cosmic web. All of these scientific prospects will start to be achievable within the next few years (see \S \ref{sec:exp}), with the SKA and SPHEREx providing the largest datasets with high sensitivity in less than a decade. 

But IM is not free of challenges as any other astronomical observation. One concerns with the proper modelling of galactic emission lines from the physical properties of the galaxy and the gas within. Observationally calibrated scaling relationships are a fast mean to model the signal but they lack the physical insight of the time consuming numerical simulations of the gas. Also, there is intrinsic variability in galactic properties but these average out for large enough voxels. A second and not less important challenge concerns the systematics that go beyond instrumental effects. Galaxy surveys either identify a source belonging to the sample or not (even if in photometric surveys stars can be misidentified), with a more or less certainty on the redshift of the source. In IM one has to disentangle the target line signal from the contaminants either using cleaning methods, component separation or cross-correlations.  

This communication was intended to give an overview of the field of Intensity Mapping as a new window into the cosmic web in the post-EoR universe. IM was first devised as a probe of the re-ionisation history of the universe, both mapping the ionised bubbles around galaxies and the global HI signal as a function of the redshift. There is an extensive literature on using the 21cm line in the EoR as well as other atomic and molecular lines, but such applications are beyond the scope of this overview. For a more thorough review in IM, all its applications and recent advances, we invite the reader to look at \emph{Line-Intensity Mapping: 2017 Status Report} \citep{Kovetz:2017agg}.

\section*{Acknowledgement}
We thank the organisers of the IAU Symposium 355: The Realm of the Low Surface Brightness Universe for the invited talk. We thank Ely Kovetz, Marta Silva, Tzu-Ching Chang, Yan Gong and Alkistis Pourtsidou for their courtesy.  JF is supported by the University of Padova under the STARS Grants programme CoGITO: Cosmology beyond Gaussianity, Inference, Theory and Observations.

\end{document}